\documentclass[aps,pra,twocolumn,superscriptaddress,showpacs]{revtex4-2}

\usepackage{amsmath,amssymb,wasysym,graphicx}
\usepackage[utf8]{inputenc}
\usepackage[T1]{fontenc}
\usepackage{xcolor}

\IfFileExists{newtxtext.sty}
{\usepackage{newtxtext,newtxmath}}
{\IfFileExists{stix.sty}
	{\usepackage{stix}}
	{\IfFileExists{mathptmx.sty}
		{\usepackage{mathptmx}}{} } }

\usepackage{textcomp}

\usepackage{bm}

\IfFileExists{siunitx.sty}{\usepackage{booktabs,siunitx}}{}

\usepackage{color}
\definecolor{LinkColor}{rgb}{0.256,0.439,0.588}
\usepackage{hyperref}
\hypersetup{
	pdfauthor={good guys},
	pdftitle={good title},
	colorlinks=true,
	citecolor=LinkColor,
	linkcolor=LinkColor,
	urlcolor=LinkColor
}

\newcommand{\beq} {\begin{equation}}
\newcommand{\eeq} {\end{equation}}
\newcommand{\bea} {\begin{eqnarray}}
\newcommand{\eea} {\end{eqnarray}}
\newcommand{\be} {\begin{equation}}
\newcommand{\ee} {\end{equation}}

\newcommand{\ket}[1]{\left|#1\right>}

\def\Fig#1{Fig.~\ref{#1}}

\begin{document}
\title{Topological edge modes and phase transition in the critical fermionic chain with long-range interaction}

\author{Wen-Hao Zhong}
\altaffiliation{The first two authors contributed equally.}
\affiliation{Department of Physics, Fuzhou University, Fuzhou 350116, Fujian, China}

\author{Wei-Lin Li}
\altaffiliation{The first two authors contributed equally.}
\affiliation {Key Laboratory of Atomic and Subatomic Structure and Quantum Control (Ministry of Education), Guangdong Basic Research Center of Excellence for Structure and Fundamental Interactions of Matter, School of Physics, South China Normal University, Guangzhou 510006, China}
\affiliation {Guangdong Provincial Key Laboratory of Quantum Engineering and Quantum Materials, Guangdong-Hong Kong Joint Laboratory of Quantum Matter, Frontier Research Institute for Physics, South China Normal University, Guangzhou 510006, China}

\author{Yong-Chang Chen}
\affiliation{Department of Physics, Fuzhou University, Fuzhou 350116, Fujian, China}

\author{Xue-Jia Yu}
\email{xuejiayu@fzu.edu.cn}
\affiliation{Department of Physics, Fuzhou University, Fuzhou 350116, Fujian, China}
\affiliation{Fujian Key Laboratory of Quantum Information and Quantum Optics,
College of Physics and Information Engineering,
Fuzhou University, Fuzhou, Fujian 350108, China}

\date{\today}

\begin{abstract}
The long-range interaction can fundamentally alter properties in gapped topological phases such as emergent massive edge modes. However, recent research has shifted attention to topological nontrivial critical points or phases, and it is natural to explore how long-range interaction influences them. In this work, we investigate the topological behavior and phase transition of extended Kitaev chains with long-range interactions, which can be derived from the critical Ising model via the Jordan-Wigner transformation in the short-range limit. Specifically, we analytically find the critical edge modes at the critical point remain stable against long-range interaction. More importantly, we observe these critical edge modes remain massless even when long-range interactions become substantially strong. As a byproduct, we numerically find that the critical behavior of the long-range model belongs to the free Majorana fermion universality class, which is entirely different from the long-range universality class in usual long-range spin models. Our work could shed new light on the interplay between long-range interactions (frustrated) and the gapless topological phases of matter.

\end{abstract}

\maketitle

\section{INTRODUCTION}
\label{sec:introduction}
For the past few decades, our understanding has primarily focused on the universal behavior of systems with short-range (SR) interactions, such as lattice systems with nearest-neighbor couplings~\cite{sachdev_2011,sachdev2023quantum,cardy1996scaling,fradkin2013field}. In recent years, there has been an investigation into the overall picture of novel phenomena in classical and quantum systems with long-range (LR) interactions~\cite{defenu2023rmp,Maity_2020}. Increasing the range of interactions $V \sim 1/r^{\alpha}$ (or equivalently reducing the power exponent $\alpha$) can fundamentally alter the "basic laws" of statistical and condensed matter physics, including the failure of quantum-classical correspondence~\cite{defenu2017prb,defenu2015pre,codello2015prd}, breakdown of the Mermin-Wagner theorem~\cite{bruno2001prl,maghrebi2017prl}, emergence of massive edge modes~\cite{vodola2014prl,Vodola_2016,viyuela2016prb,lepori2016effective}, and new LR universality classes~\cite{fisher1972prl,brezin2014crossover,angelini2014pre,behan2017prl,Behan_2017,yu2022prb,yu2023pre,yu2023prb}, among others. Moreover, the experimental realization of quantum systems with LR interactions, such as cold atomic gases in cavities~\cite{bloch2008rmp,ritsch2013rmp,mivehvar2021cavity}, trapped ions~\cite{ladd2010quantum,monroe2021rmp}, as well as programmable Rydberg quantum simulators~\cite{bernien2017probing,ebadi2021quantum,keesling2019quantum,semeghini2021probing}, has spurred significant motivation to explore the novel properties of such systems.

On a different front, the development of topological phases of matter has garnered significant attention in the past few decades, expanding our understanding of quantum matter beyond the Landau paradigm~\cite{chen2012symmetry,wen2019choreographed,wen2017rmp,chen2013prb,senthil2015symmetry,yu2024prl}. A notable example is the gapped topological insulator and superconductors~\cite{hasan2010rmp,qi2011rmp}, where the bulk is gapped and nontrivial gapless edge modes emerge at the boundary~\cite{witten2016three,pollmann2010prb,fidkowski2011prb,pollmann2012prb}. However, recent research has revealed that many key features of topological physics, such as topological invariant~\cite{ruben2021prx,yu2022prl,parker2018prb,verresen2020topology}, universal bulk-boundary correspondence in the entanglement spectrum~\cite{yu2024universal}, and degenerate edge modes~\cite{hidaka2022prb,su2024gapless,wen2023prb,duque2021prb,ando2024gauging,prembabu2022boundary,umberto2021scipost,huang2023topological,chang2022absence,yu2024quantum,wen2023classification,scaffidi2017prx,thorngren2021prb,li2023decorated,li2023intrinsicallypurely,nathanan2023scipost,nathanan2023scipost_b}, persist despite nontrivial coupling between the boundary and critical bulk modes. Nevertheless, there are still largely unexplored areas within the field of the interplay between topology and quantum criticality even in the free fermion system~\cite{ruben2018prl,verresen2020topology,Jones_2019,choi2023finite,kumar2023prb,kumar2021multi}.

Recently, there has been a surge of interest in the LR version of an integrable topological fermionic chain, featuring a LR superconducting pairing or hopping term~\cite{dutta2017prb,van2016pra,francia2022prb,jager2020prb,mondal2022prb,alecce2017prb,ares2018pra,fraxanet2021prr,gong2016prb,patrick2017prl}, which exhibits exotic properties such as massive edge modes~\cite{vodola2014prl,Vodola_2016,viyuela2016prb,lepori2016effective}, anomalous behavior of correlation functions~\cite{Vodola_2016}, and novel bulk-boundary correspondence~\cite{jones2023prl}. Unlike the SR case, the LR fermionic and spin chains are two independent models that cannot be mapped to each other through the Jordan-Wigner transformation. Furthermore, LR power-law interactions are ubiquitous in modern quantum simulation platforms~\cite{bloch2008rmp,ladd2010quantum,bernien2017probing}. Therefore, it is fundamentally important to study LR interacting models both theoretically and experimentally. However, over the past few decades, most research efforts have been focused on the stability of edge modes in gapped topological phases with LR interactions~\cite{vodola2014prl,Vodola_2016,viyuela2016prb,lepori2016effective,viyuela2018prl}. For gapless topological phases, such as topologically nontrivial quantum critical points (QCPs), it remains unclear how the critical edge modes remain stable against LR interactions. Moreover, can LR interactions induce a crossover of the critical edge modes from massless to massive, similar to the effects observed in LR interacting gapped topological phases?

To answer the series of questions outlined above, in this work, we investigate the topological and critical behavior in the extended Kitaev chains with LR interactions, which can be derived from the cluster Ising model via the Jordan-Wigner transformation in the SR limit. Specifically, through exact analytical calculations, we find edge modes at the critical point between distinct topological superconductors (TSC) that remain stable under LR interactions. Furthermore, we observe that the edge modes remain massless even when LR interactions become substantial strong, a scenario entirely different from the gapped topological phases with LR interactions. As a byproduct, we determine the critical exponent of the LR interacting model always consistent with the universality class of the SR one (free Majorana fermion) through finite-size scaling.

The paper is organized as follows: Section~\ref{sec:model} introduces the lattice model of the extended Kitaev chain with LR interactions. Section~\ref{sec:phase} presents the global phase diagram of the model, finite-size scaling for fidelity susceptibility, and edge modes at the critical points with LR interactions. The conclusion is provided in Sec.~\ref{sec:con}. Additional data for analytical and numerical calculations are included in the Appendix.

\section{MODEL AND METHOD}%
\label{sec:model}
\subsection{Extended Kitaev chain with long-range interaction}
In this work, we investigate the behavior of LR interacting fermionic particles on a lattice of length $L$ (see Fig.~\ref{fig1}(a)). The model $H_{\text{LR}}$ is described by the following Hamiltonian:
\begin{equation}
\label{E1}
\begin{split}
&H_{\text{LR}} = -\Delta\sum_{j=1}^{L} \sum_{l=1}^{L-1} d_{l}^{-\alpha} (c_j^\dagger c_{j+l} + c_j^\dagger c_{j+l}^\dagger+ h.c.) \\
& - h\sum_{j=1}^{L-2} (c_j^\dagger c_{j+2} + c_j^\dagger c_{j+2}^\dagger + h.c.).
\end{split}
\end{equation}
Here, $c^{\dagger}_{j}$ ($c_{j}$) represents the fermionic creation (annihilation) operator at site $j$. The parameter $h$ denotes the strength of the next nearest neighbor fermion $p$-wave pairing and hopping amplitude. For a periodic chain, we define $d_{l}=l$ ($d_l = L-l$) if $l<L/2$ ($l > L/2$), and choose antiperiodic boundary conditions. For an open chain, we define $d_l=l$ and omit terms involving $c_{j>L}$. In this work, we set $\Delta=1$ as energy unit.

The Hamiltonian~(\ref{E1}) in the SR limit can be derived from the cluster Ising model through the Jordan-Wigner transformation. The latter is an extension of the SR Ising model and falls within the topologically nontrivial Ising universality class (symmetry-enriched Ising criticality)~\cite{ruben2018prl,ruben2021prx,yu2022prl,yu2024quantum}, exhibiting gapped spontaneous symmetry breaking (SSB) and cluster symmetry-protected topological (SPT) phase for $h<1.0$ and $h>1.0$, respectively, separated by a topologically nontrivial critical point at $h=1.0$~\cite{Son_2011,smacchia2011pra,guo2022pra,verresen2017prb}. Upon the Jordan-Wigner transformation, the SSB and SPT phases map to the TSC phase with winding numbers of $1$ and $2$, respectively. The phase transition between them is characterized by topologically protected Dirac cones~\cite{verresen2020topology}.

However, with LR interactions, the Hamiltonian~(\ref{E1}) is no longer connected to the LR interacting cluster Ising model by a Jordan-Wigner transformation, indicating that their respective phase diagrams may differ. Fortunately, unlike the LR cluster Ising model, Eq.~(\ref{E1}) remains exactly solvable, enabling analytical solutions at any finite $\alpha$. In the following, we explore the topological behavior and phase transition of the LR fermionic model by examining its winding number, and fidelity susceptibility for the periodic chain, as well as  energy spectrum and the critical edge modes for the open one.

\section{PHASE DIAGRAM AND CRITICAL BEHAVIOR}
\label{sec:phase}

\subsection{Quantum phase diagram}
\begin{figure}
   \includegraphics[width=8.6cm]{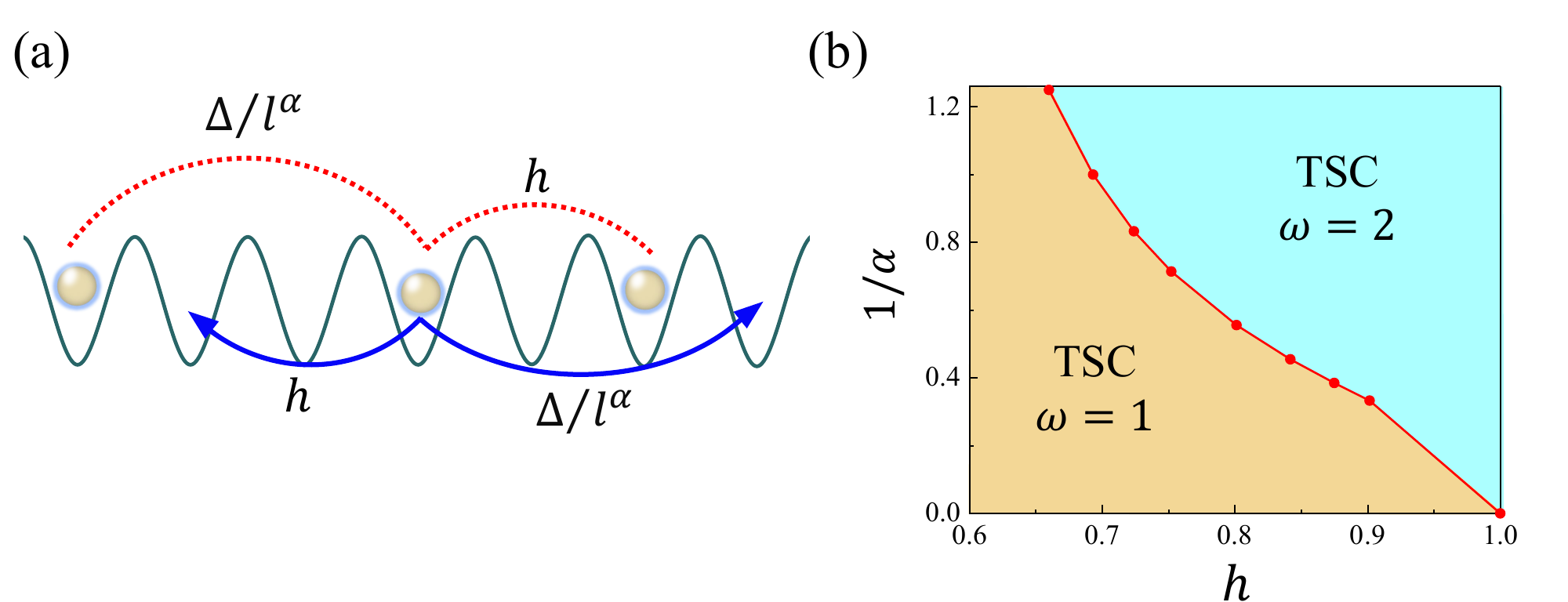}
    \caption{
    (Color online) (a) A schematic representation of the LR interacting fermionic chain is presented. The brown ball denotes the fermion, while the blue arrow line (red dashed line) represents the LR and next-nearest-neighbor hopping (pairing) term. (b) The global phase diagram illustrates the extended Kitaev chain with LR interaction as a function of the LR power exponent $\alpha$ and the driven parameter $h$. The diagram delineates the topological superconductor with winding number $\omega$, denoted as the TSC ($\omega=1$) phase (brown region), and the TSC ($\omega=2$) phase (light blue region). Red points mark the critical points $h_c^*$ corresponding to different $\alpha$, and the red line forms the critical line between the two phases.}
    \label{fig1}
\end{figure}

Thanks to the integrability of the model, Eq.~(\ref{E1}) can be transformed into momentum space using Fourier transformation:
\begin{equation}
\label{E2}
H_{\text{LR}} = \sum_k [iy_k (c_k c_{-k} + c_k^\dagger c_{-k}^\dagger) + z_k (c_k^\dagger c_k + c_{-k}^\dagger c_{-k}-1)].
\end{equation}
Here, the wave vector $k$ belongs to $\pm\{2n-1\}\pi/L,n=1,2,...L/2$, and $y_k = [-h \sin(2k) - \Delta f_\alpha(k)]$, $z_k = [-h \cos(2k) - \Delta g_\alpha(k)]$ with $f_\alpha(k) = \sum_{l} \sin(kl)/d_l^\alpha $ and $g_\alpha(k) = \sum_{l} \cos(kl)/d_l^\alpha $. The Hamiltonian is already in a small Hilbert space and can be easily diagonalized by the Bogoliubov transformation(see Appendix~\ref{sec:appA} for the details of analytical calculations), and the ground state is given by: $|G \rangle = \prod_{k>0}[{\rm{cos}}(\frac{\theta_{k}}{2})+\mathrm{i}{\rm{sin}}(\frac{\theta_{k}}{2})c^{\dagger}_{k}c^{\dagger}_{-k}]|{\rm{Vac}}\rangle$ ($|{\rm{Vac}} \rangle$ is the vacuum state of $c$ fermions).

Before delving into the details of analytical results, let's summarize our main findings and outline the global quantum phase diagram of the model in Eq.~(\ref{E1}). The schematic phase diagram is provided in Fig.~\ref{fig1}(b). As $\alpha$ approaches infinity, the model can be derived from the cluster Ising model via Jordan-Wigner duality, and the TSC phases transform to SSB or SPT phases, depending on the winding number $\omega=1$ (brown region) or $2$ (light blue region). More importantly, the phase transition between them exhibits a topologically protected Dirac cone with critical edge mode in open boundary. When $\alpha$ is finite, we observe that both TSC phases with different winding numbers are stable against LR interactions. Red points mark the critical points corresponding to different $\alpha$, and the red line forms the critical line between the two phases. Whether LR interactions influence the physics of critical edge modes is the crucial issue in this work, and we will address it in the following subsection.

\subsection{Topological phases with LR interactions}

\begin{figure}
\includegraphics[width=8.6cm]{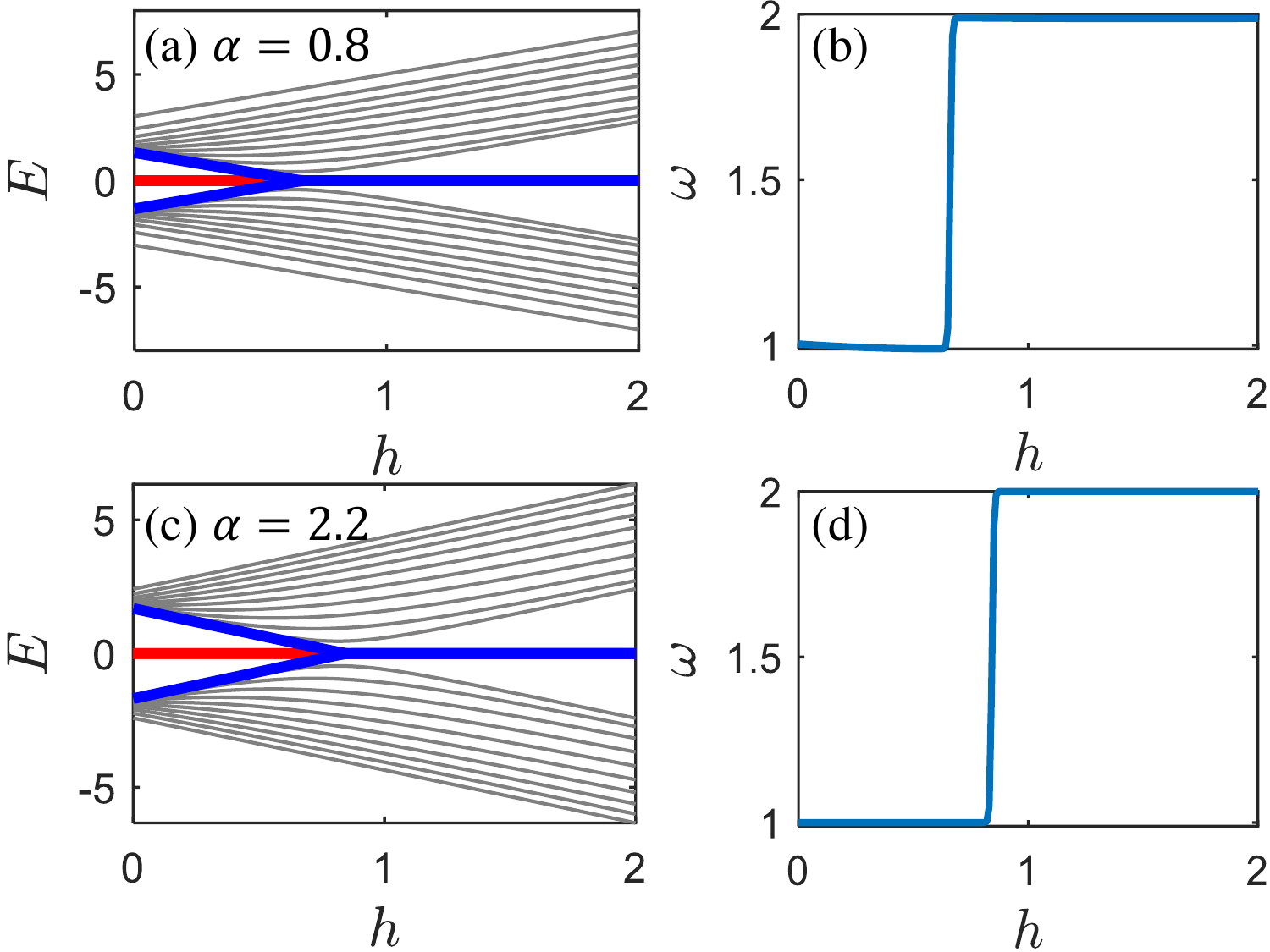}
\caption{(Color online) (a) The energy spectrum as a function of $h$ for the extended Kitaev chain with LR interactions with $\alpha=0.8$ (a) and $2.2$ (c). (b) The winding number as a function of $h$ for the extended Kitaev chain with LR interactions with $\alpha=0.8$ (b) and $2.2$(d). The energy spectrum calculations are conducted under open boundary conditions, while the winding number calculations are performed under periodic boundary conditions. The simulated system size is $L=240$.}
\label{fig2}
\end{figure}


In this section, we first identify the possible topological phases in the LR model. Since the gapped topological phases exhibit degenerate edge modes in open boundary conditions (OBCs), we separately calculate the energy spectrum and winding number as functions of $h$ to explicitly demonstrate the possible edge modes at the boundary.

To acquire the winding number, following Ref.~\cite{zhang2015prl}, we can reformulate the Hamiltonian as follows:
\begin{equation}
H_{\text{LR}} = 4\sum_{k>0} \mathbf{r}_k \cdot \mathbf{s}_k,
\end{equation}
where $\mathbf{r}_k = (0,y_k,z_k)$ and the pseudospin $\mathbf{s}_k = \frac{1}{2}[c_{-k}^\dagger c_k^\dagger - c_{-k} c_k, i(c_k c_{-k} + c_k^\dagger c_{-k}^\dagger), c_k^\dagger c_k + c_{-k}^\dagger c_{-k}-1]$. These pseudospin operators satisfy the $SU(2)$ algebra, and the winding number in the auxiliary space $(y,z)$ is defined as:
\begin{equation}
\omega = \frac{1}{2\pi} \oint \frac{1}{r^2} (zdy - ydz),
\end{equation}
where the integral spans over the space $(y,z)$, covering all $k$ values from $0$ to $2\pi$. $\omega$ serves as a means to distinguish different topological phases, which possess different winding numbers.

To be precise, we calculated the energy spectrum as a function of $h$ for $\alpha > 1$ ($\alpha=2.2$) and $\alpha < 1$ ($\alpha=0.8$) under OBCs, respectively. As depicted in Fig~\ref{fig2} (a) and (c), we find that when $h$ is less than the critical point, denoted by $h_c$, there is only a pair of zero-energy edge modes in the energy spectrum (indicated by the red line in the Fig.~\ref{fig2}). Conversely, when $h>h_c$, there are two pairs of zero-energy edge modes in the energy spectrum (indicated by blue line in the Fig.~\ref{fig2}). Furthermore, we calculated the winding number as a function of $h$ for the same $\alpha=0.8$ and $2.2$, as shown in Fig.~\ref{fig2}(b) and (d). Before and after the phase transition point $h_c$, the winding number changes from $1$ to $2$, indicating that the number of edge modes changes from one to two pairs, which is consistent with the observations of the energy spectrum(see Appendix~\ref{sec:appB} for other interaction powers $\alpha$). These findings strongly suggest that even with relatively strong LR interactions, the TSC phases of different winding numbers remain stable. In the following subsection, we will explore the quantum critical behavior between distinct TSC phases and the topological properties of underlying critical points.

\subsection{Finite-size scaling and critical exponents}
\begin{figure}
\includegraphics[width=8.6cm]{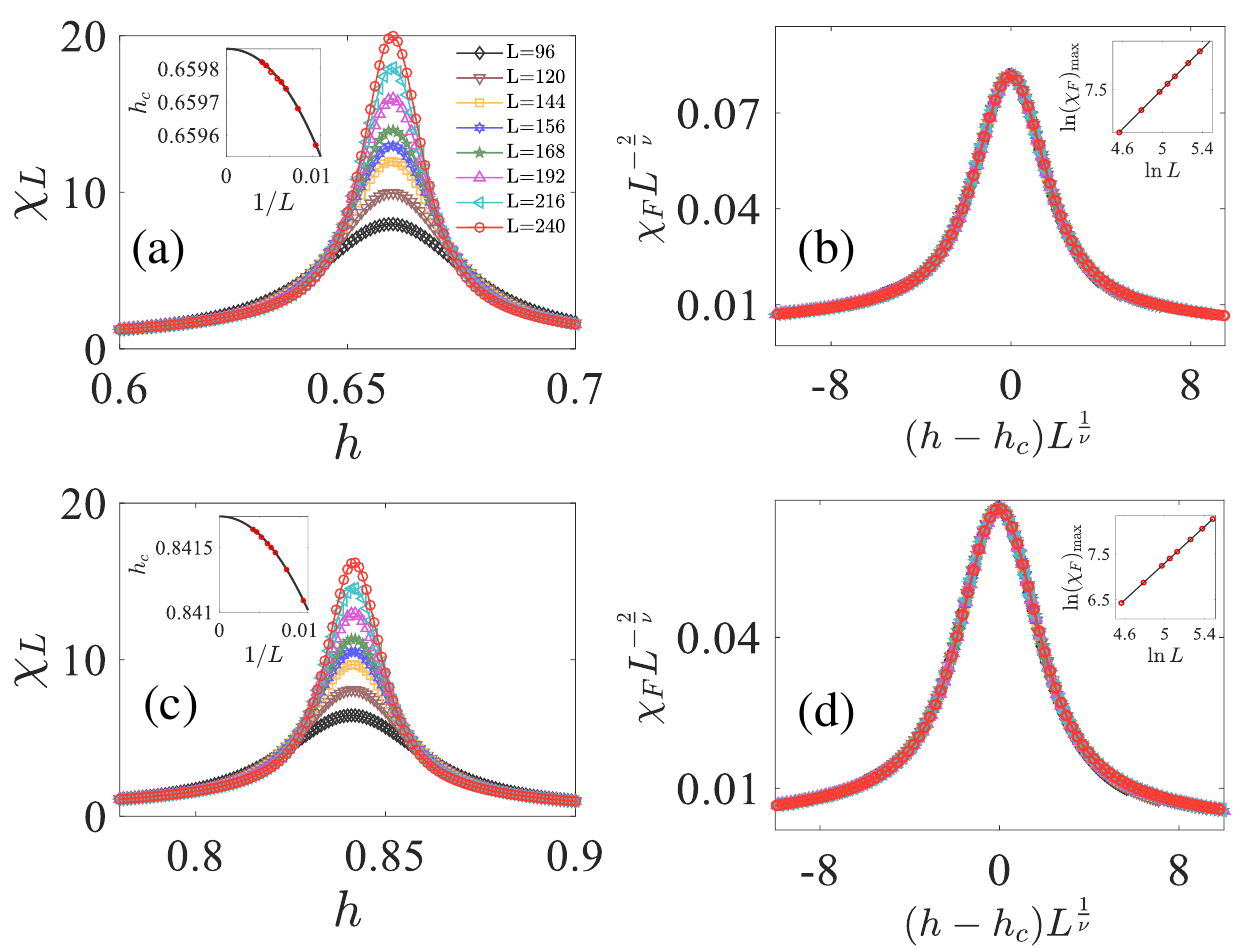}
\caption{(Color online) Fidelity susceptibility per site $\chi_{L}$ of the extended Kitaev chain with LR interaction for $\alpha=0.8$ (a) and $2.2$ (c) with $L=96,120,144,156,168,192,216,240$ sites as a function of $h$; The insert exhibits the extrapolation of critical point $h_{c}^{*}$ for the model. We use polynomial fitting $h_{c}(L) = h_{c}^{*} + aL^{-1/\nu}$ and extrapolate the critical point $h_{c}^{*}$ = 0.65986(1) (0.84174(1)) for $\alpha=0.8$ ($2.2$). Data collapse of fidelity susceptibility $\chi_{F}$ for the extended Kitaev chain with $\alpha=0.8$ (b) and $2.2$ (d), where $\nu=0.9983(2)$ (0.9966(2)) and $h_{c}^{*}$ = 0.65986(1) (0.84174(1)) are used for data collapse plots. The insert shows the maximal fidelity susceptibility per site $\chi_{L}=\chi_{F}/L$ as a function of system sizes $L$ for $\alpha=0.8$ (b) and (d) $2.2$ . We use polynomial fitting $\chi_{F}(h_{c}(L))=L^{\mu}(c+dL^{-1})$ and extrapolate the critical adiabatic dimension $\mu$ = 2.0034(3) (2.0068(4)) for $\alpha=0.8$ ($2.2$).}  
\label{fig3}
\end{figure} 

After delineating all the quantum phases in the phase diagram, we shift our attention to the more intriguing QPTs between these phases and inquire about the scaling and critical exponents at these phase transition points. As mentioned above, the phase transition between topologically distinct superconducting phases belongs to a topological nontrivial universality class and exhibits robust edge modes in the open boundary. We will investigate whether LR interactions influence the physics of critical edge modes in the next subsection. In this section, we determine the critical points and critical exponents for various LR power exponents $\alpha$ through the finite-size scaling of fidelity susceptibility.

The concept of fidelity susceptibility pertains to a system undergoing a continuous phase transition from an ordered to a disordered phase upon tuning the parameter $h$ to a critical value $h_{c}$. At this point, the structure of the ground state wave function changes significantly. The quantum ground-state fidelity $F(h,h + \delta h)$ quantifies the overlapping amplitude between the ground-state wave function at external field $h$ and $h+\delta h$. Near $h_{c}$, $F(h_{c},h_{c} + \delta h)\sim 0$, indicating a drastic change in the ground state. Then, the fidelity susceptibility, defined as the leading term in the fidelity~\cite{gu2010fidelity,Gu_2009}:

\begin{equation}
\begin{split}
\label{E2}
\chi_{F}(h)=\lim_{\delta h \rightarrow 0}\frac{2(1-F(h,h+\delta h))}{(\delta h)^{2}} = \frac{1}{4}\sum_{k>0}\left(\frac{\partial\theta_{k}(h)}{\partial h}\right)^2.
\end{split}
\end{equation}

For a continuous QPT in a finite system size $L$, the fidelity susceptibility $\chi_{F}(h)$ exhibits a peak at a critical point $h_{c}(L)$, and the value of the quantum critical point $h_{c}^{*}$ can be estimated by polynomial fitting $h_{c}(L)=h_{c}^{*}+aL^{-1/\nu}$~\cite{sandvik2010computational}. In the vicinity of $h_{c}(L)$, previous research~\cite{schwandt2009prl,albuquerque2010prb,konig2016prb,zhang2023critical,yu2023pra} have shown that the finite-size scaling behaviors
of fidelity susceptibility $\chi_{F}(h)$ follows:
\begin{equation}
\label{Emu}
\begin{split}
&\chi_{F}(h\rightarrow h_{c}(L))\propto L^{\mu} \\
\end{split}
\end{equation}
and
\begin{equation}
\begin{split}
\label{fs_scaling}
L^{-d}\chi_{F}(h) = L^{(2/\nu)-d}f_{\chi_{F}}(L^{1/\nu}|h-h_{c}|),
\end{split}
\end{equation}
where $\mu (=2+2z-2 \Delta_{V} )$ is the critical adiabatic dimension~\cite{gu2010fidelity}. $\Delta_{V}$ is the scaling dimension of the local interaction $V(x)$ at $h_{c}^{*}$,  $\nu$ is correlation length exponent, and it can be easily computed according to the relation: $\nu=2/\mu$. $z$ is dynamic exponent, $d$ is the spatial dimension of the system, and $f_{\chi_{F}}$ is an unknown scaling function.  Based on Eq.~(\ref{Emu}) and~(\ref{E2}), the values of critical exponents $\nu$ and $\mu$ of the QPT can be determined, and the critical behavior of the LR interacting quantum system can be easily determined. Note that in practice, the critical exponent $\mu$ is usually extracted from fidelity susceptibility per site, $\chi_{L}(f) = \chi_{F}(h)/L^{d}$.

The finite-size scaling behavior of fidelity susceptibility for $\alpha = 0.8$ and $2.2$ with different $L$ is presented in Fig.\ref{fig3}(a) and (c), which obeys $\chi_{L}(h_{c}(L))\propto L^{\mu-1}$ (Eq.(\ref{Emu})) near the second-order QPT critical point. As the system size $L$ increases, the peak position $h_{c}(L)$ approaches the exact critical point $h_{c}^{*}$, indicating the stability of the critical points regardless of the LR interaction strength. More precisely, for the LR interacting extended Kitaev chain with $\alpha=0.8$ and $2.2$, $h_{c}^{*}$ is determined by polynomial fitting $h_{c}(L) = h_{c}^{*} + aL^{-1/\nu} $, and then extrapolating to $L$ to infinity (see insert in Fig.~\ref{fig3}(a) and (c)). According to Eq.~(\ref{fs_scaling}), the fidelity susceptibility follows an exact scaling relation and collapses to a single master curve (Fig.~\ref{fig3}(b) and (d)), confirming the correctness of the extrapolation. The finite-size scaling behavior of fidelity susceptibility for other $\alpha$ values is also investigated (see Appendix~\ref{sec:appD} for details), and the results are presented in Table~\ref{tab:exponents}. Results show that the QCP shifts to higher $h_{c}^{*}$ values as $\alpha$ increases.

The next key questions pertaining to the critical behavior of LR interacting extended Kitaev chains for different $\alpha$ and whether there exists a critical threshold $\alpha_c$ at which the critical behavior transitions continuously from a LR universality class to a SR one. To address these questions, we calculate the critical exponents $\mu$ and $\nu$ of the LR model in the region $0.8 \leq \alpha \leq 3.0$ for various system sizes $L$. The fidelity susceptibility per site, $\chi_{L}=\chi_{F}/L$, at the peak position $h_{c}(L)$ for substantially strong ($\alpha=0.8$) and small ($\alpha=2.2$) LR interactions are depicted in Fig.\ref{fig3}(a) and (c) (see Appendix~\ref{sec:appC} for other LR power exponents $\alpha$). The adiabatic critical dimension $\mu$ can be accurately fitted using a polynomial function of $\chi_{L}(h_{c}(L))=L^{\mu-1}(c+dL^{-1})$, as demonstrated in the inset of Fig.~\ref{fig3}(b) and (d).

\begin{figure}
\includegraphics[width=8.6cm]{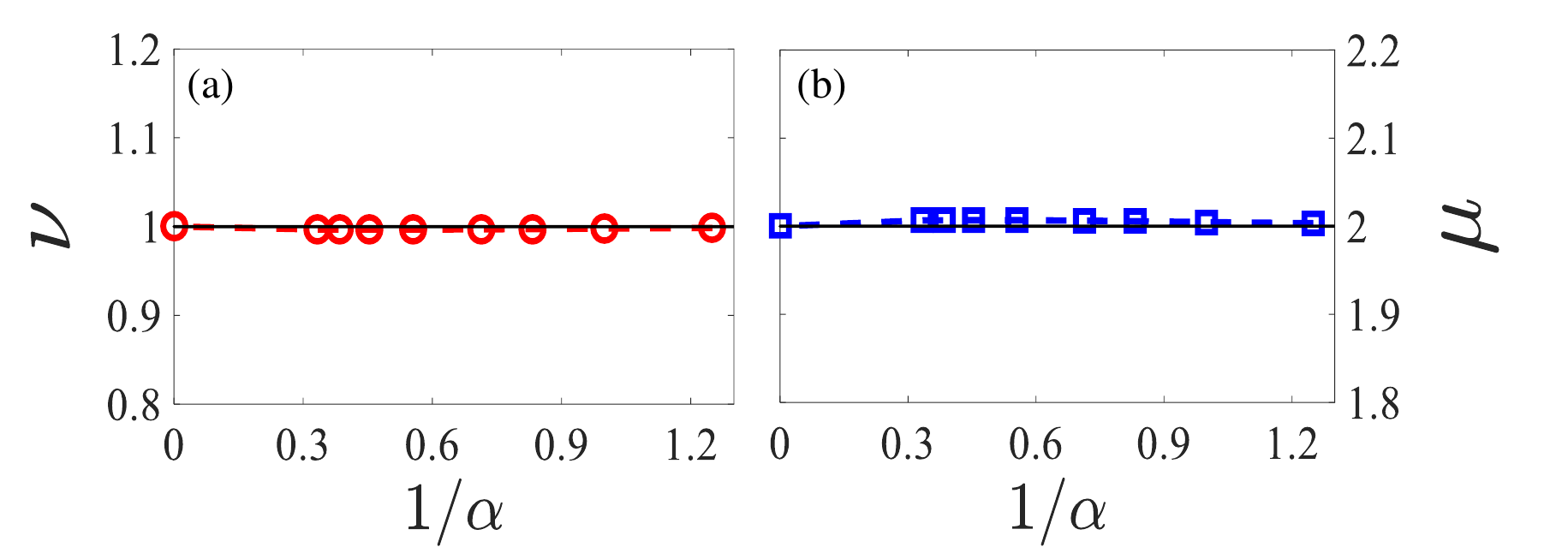}
\caption{(Color online) Critical exponent of the correlation length $\nu$ (black dash line refers to SR Kitaev chain correlation length exponent $\nu=1.0$ as a comparison) (a) and critical adiabatic dimension $\mu$ (black dash line refers to SR Kitaev chain critical adiabatic dimension $\mu=2.0$ as a comparison) (b) concerning $1/\alpha$ for the extended Kitaev chain with LR interaction.}
\label{fig4}
\end{figure}

Following Eq.~(\ref{fs_scaling}), the fidelity susceptibility can be scaled by $L^{-2/\nu}\chi_{F}$ as a function of $L^{1/\nu}(h-h_{c}^{*})$ in the vicinity of the QCP $h_{c}^{*}$. The critical correlation length exponent $\nu$ is then determined by $\nu=2/\mu$. Substituting the obtained critical point $h_{c}^{*}$ and critical exponent $\nu$ into Eq.~(\ref{fs_scaling}), all fidelity susceptibilities for different $L$ collapse into a single curve (Fig.~\ref{fig3}(b) and (d)), indicating the accuracy of the estimated critical point and critical exponent. It is worth noting that the peak of the data collapse is not at $0$ due to the finite-size effect arising from $h_{c}(L)=h_{c}^{*}+aL^{-1/\nu}$, where $a \neq 0$. The calculations of the critical adiabatic dimension $\mu$ and the correlation length exponent $\nu$ for other values of $\alpha$ are presented in Appendix~\ref{sec:appD}, and the results for all $\alpha$ are summarized in Table~\ref{tab:exponents}. As observed in Fig.~\ref{fig4}(a) and (b), both $\nu$ and $\mu$ as functions of $1/\alpha$ remain relatively constant and approach the critical exponent of the Kitaev chain in the SR limit, $\nu=1.0$ and $\mu=2.0$, respectively~\cite{francesco2012conformal,ginsparg1988applied}, within the $0.34\%$ error due to finite size effects (indicated by the black solid line in Fig.~\ref{fig4}). Therefore, the critical behavior of the extended Kitaev chain remains unchanged regardless of the strength of the LR interaction, which indicates LR interaction is irrelevant under the renormalization group, and consequently, long-distance properties of the LR model are not significantly modified relative to the SR ground state. We defer the renormalization group study for the LR interacting extended Kitaev chain to future work. It is worth emphasizing that the above behavior contrasts with typical LR interacting quantum spin chains, which exhibit new LR universality classes\cite{defenu2015pre,defenu2017prb,yu2023pre}.

\begin{ruledtabular}
\begin{table}[tb]
\caption{Critical exponents of the extended Kitaev chain with LR interaction  for different $\alpha$. Critical exponents in the cluster Ising chain ($\alpha=\infty$) are also listed for comparison. }
\label{tab:exponents}
\begin{tabular}{cccc}
$\alpha$   & $h_{c}^{*}$			   & $\nu$	
           & $\mu$	         \\
\hline
0.8	       & 0.65986(1)				   & 0.9983(2)			       & 2.0034(3)		\\
1.0		   & 0.69315(1)				   & 0.9977(2)			       & 2.0046(3)		\\
1.2		   & 0.72383(1)				   & 0.9969(2)	
           & 2.0056(3)	     \\
1.4		   & 0.75199(1)				   & 0.9966(2)				   & 2.0063(3)	    \\
1.8        & 0.80117(1)                    & 0.9966(2)                    & 2.0068(4)       \\
2.2        & 0.84174(1)                    & 0.9966(2)                     & 2.0068(4)       \\
2.6		   & 0.87484(1)				   & 0.9967(2)				   & 2.0067(4)       \\
3.0		   & 0.90154(1)				   & 0.9967(2)				   & 2.0066(4)	    \\
$\infty$   & 1.00000				   & 1.00000				   & 2.00000	    \\
\end{tabular}
\end{table}
\end{ruledtabular}

\subsection{Topological critical edge modes with LR interactions }
\begin{figure}[t]
\includegraphics[width=8.6cm]{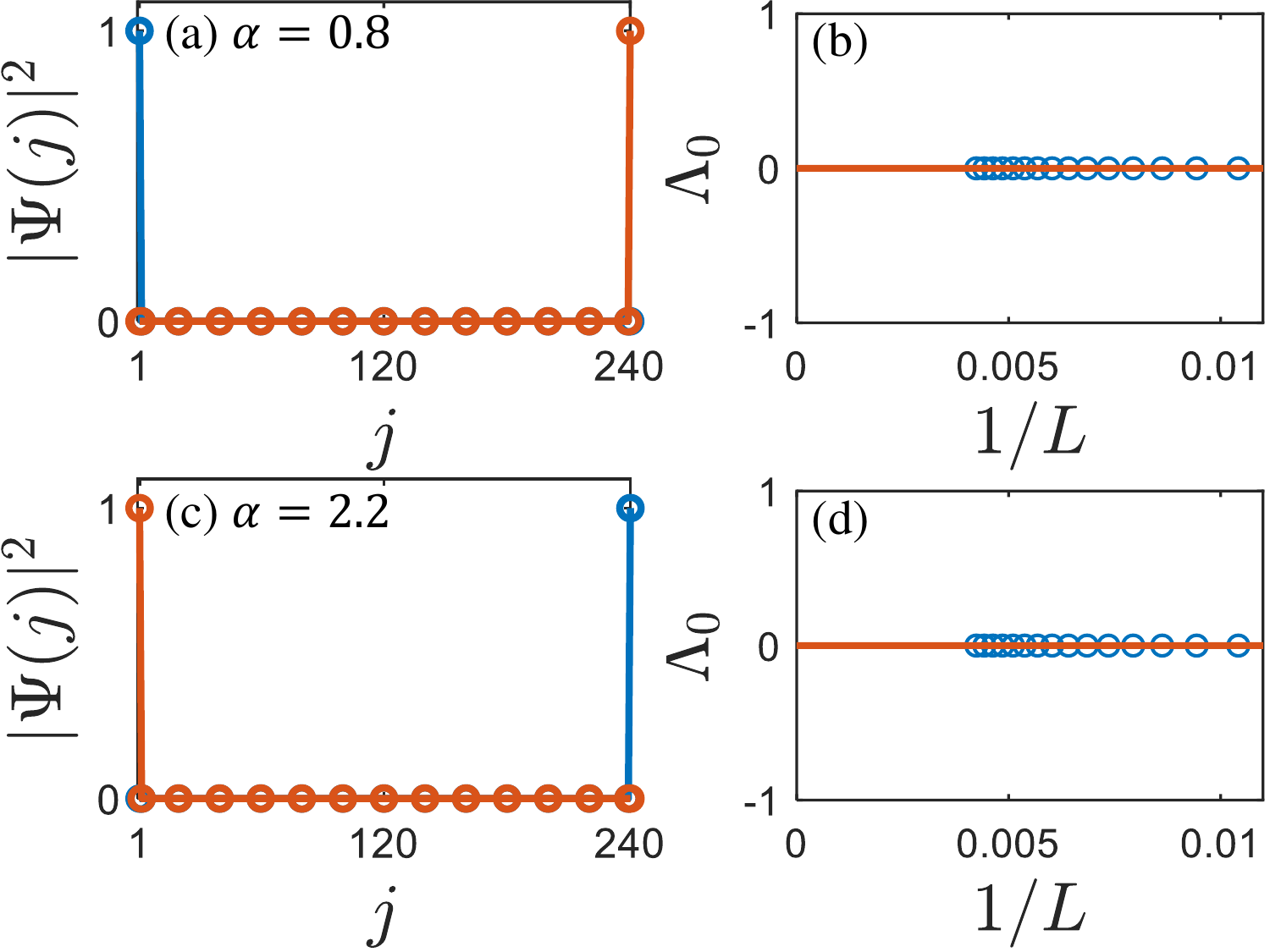}
\caption{(Color online) Wave function distributions at the critical point for $\alpha = 0.8$ (a) and $2.2$ (c). Finite size scaling of the edge mode mass $\Lambda_0$ at the critical point for $\alpha = 0.8$ (b) and $2.2$ (d). All calculations are conducted under OBCs, and the simulated system size is $L=240$.}
\label{fig5}
\end{figure} 

In the SR limit, a crucial aspect of the phase transition between topologically distinct superconducting phases is the exhibition of topologically protected edge modes, even in the presence of non-trivial coupling between the boundary and the critical bulk modes. Previous studies~\cite {vodola2014prl,viyuela2016prb} has demonstrated that the edge mode in the gapped topological phase remains stable against LR interactions, and a sufficiently strong LR interaction can convert the massless edge mode into a new massive one~\cite{Vodola_2016,lepori2016effective}. It is natural to inquire whether the edge mode at the topological nontrivial critical point is stable against LR interactions and whether a substantial strong LR interaction can induce novel massive edge modes.

To address these questions, we computed the edge mode mass $\Lambda_{0}$ and wave function distribution $|\Psi(j)|^{2}$ at the critical point for the LR model under OBCs (see Appendix~\ref{sec:AppE} for the calculation details) for both substantial strong ($\alpha=0.8$) and weak ($\alpha=2.2$) LR interactions. As depicted in Fig.\ref{fig5} (a) and (c), we observed that no matter how LR interaction is, the wave function distribution at the critical point exhibits prominent peaks at the boundary, indicating the robustness of the critical edge mode even under substantially strong LR interactions ($\alpha < 1$). More importantly, as illustrated in Fig.\ref{fig5} (b) and (d), we found that regardless of the LR interaction strength, the critical edge mode mass remains zero in the thermodynamic limit, which is completely different from the LR interaction induced massive edge mode observed in gapped topological phases. In the Appendix~\ref{sec:AppE}, we provide additional evidence regarding the critical edge mode mass and wave function distributions for various other $\alpha$ values, further confirming the robustness of the critical massless edge mode against LR interactions.


\section{CONCLUSION AND OUTLOOK}
\label{sec:con}
To summarize, we focus on the topologically protected edge modes and phase transitions in LR critical extended Kitaev chains. Utilizing the Jordan-Wigner transformation, we have confirmed the stability of edge modes at critical points between topologically distinct superconducting phases against LR interactions. More importantly, we observe that the edge modes remain massless even when the LR power exponent $\alpha$ is substantially small, which contrasts with the massive edge mode induced by LR interactions in the gapped topological phases. Additionally, as a byproduct, we have determined the critical exponents of the extended Kitaev chain with different LR powers $\alpha$ remain unchanged through finite-size scaling, aligning with the universality class of the SR Kitaev chain. Intriguing future inquiries include understanding the underlying physical mechanisms behind the notable differences between LR-interacting gapped and gapless topological phases or critical points, as well as their possible higher-dimensional generalizations. Our work could shed new light on the interplay between LR interactions (frustrated) and the gapless topological phases of matter.

\begin{acknowledgments}
This work is supported by the start-up grant XRC-23102 of Fuzhou University.

\end{acknowledgments}

\bibliography{main}

\newpage
\onecolumngrid

\appendix

\section{ANALYTICAL CALCULATIONS DETAILS}
\label{sec:appA}
In this section, we provide details of the analytic derivation presented in the main text. This derivation relies on the integrability of the LR fermionic Hamiltonian given in Eq.~(\ref{E1}), which is obtained through Fourier transformation:  $c_k = \frac{1}{\sqrt{L}} \sum_{j=1}^{L} c_j e^{-ikj}$, where $k = \pm\frac{(2n-1) \pi }{L}$, $n = 0, 1, 2, \ldots, \frac{L}{2}$,
\begin{equation}
H_{\text{LR}} = \sum_k [iy_k (c_k c_{-k} + c_k^\dagger c_{-k}^\dagger) + z_k (c_k^\dagger c_k + c_{-k}^\dagger c_{-k}-1)].
\end{equation}
Here, $y_k = -h \sin(2k) - \Delta f_\alpha(k)$ and $z_k = -h \cos(2k) - \Delta g_\alpha(k)$ with $f_\alpha(k) = \sum_{l} \sin(kl)/d_l^\alpha $ and $g_\alpha(k) = \sum_{l} \cos(kl)/d_l^\alpha $.
Using the Bogoliubov transformation, defined as:
\begin{align}
\label{E_gamma}
\gamma_k &= \cos(\frac{\theta_k}{2}) c_k - i\sin(\frac{\theta_k}{2}) c_{-k}^\dagger 
\end{align}
\begin{align}
\label{E_gamma^dagger}
\gamma_k^\dagger &= \cos(\frac{\theta_k}{2}) c_k^\dagger + i\sin(\frac{\theta_k}{2}) c_{-k},
\end{align}
with
\begin{align}
\label{E5}
\tan(\theta_k) = -\frac{y_k}{z_k},
\end{align}
the Hamiltonian can be diagonalized as:
\begin{equation}
H_{\text{LR}} = \sum_{k>0} \epsilon_k (\gamma_k^\dagger \gamma_k - \frac{1}{2}),
\end{equation}
where 
\begin{equation}
\label{E4}
\epsilon_k = 4\sqrt{y_k^2 + z_k^2}.
\end{equation}
Finally, the ground state $|G\rangle$ of the model is given by: 
\begin{equation}
|G\rangle = \prod_{k>0} [\cos(\frac{\theta_k}{2}) + i\sin(\frac{\theta_k}{2}) c_k^\dagger c_{-k}^\dagger] \rm |Vac\rangle,
\end{equation}
where $\rm |Vac\rangle$ denotes the vacuum state of the $c$ fermion.

Once the ground state and energy spectrum of the model are obtained, we can calculate different types of physical quantities to study the various phases and phase transitions in the phase diagram.

\section{ENERGY SPECTRUM AND WINDING NUMBER FOR OTHER INTERACTION POWERS}
\label{sec:appB}
In this section, we provide additional data demonstrating the energy spectrum under OBCs and the winding number for other interaction powers $\alpha$.

\begin{figure}
\includegraphics[width=0.8\textwidth]{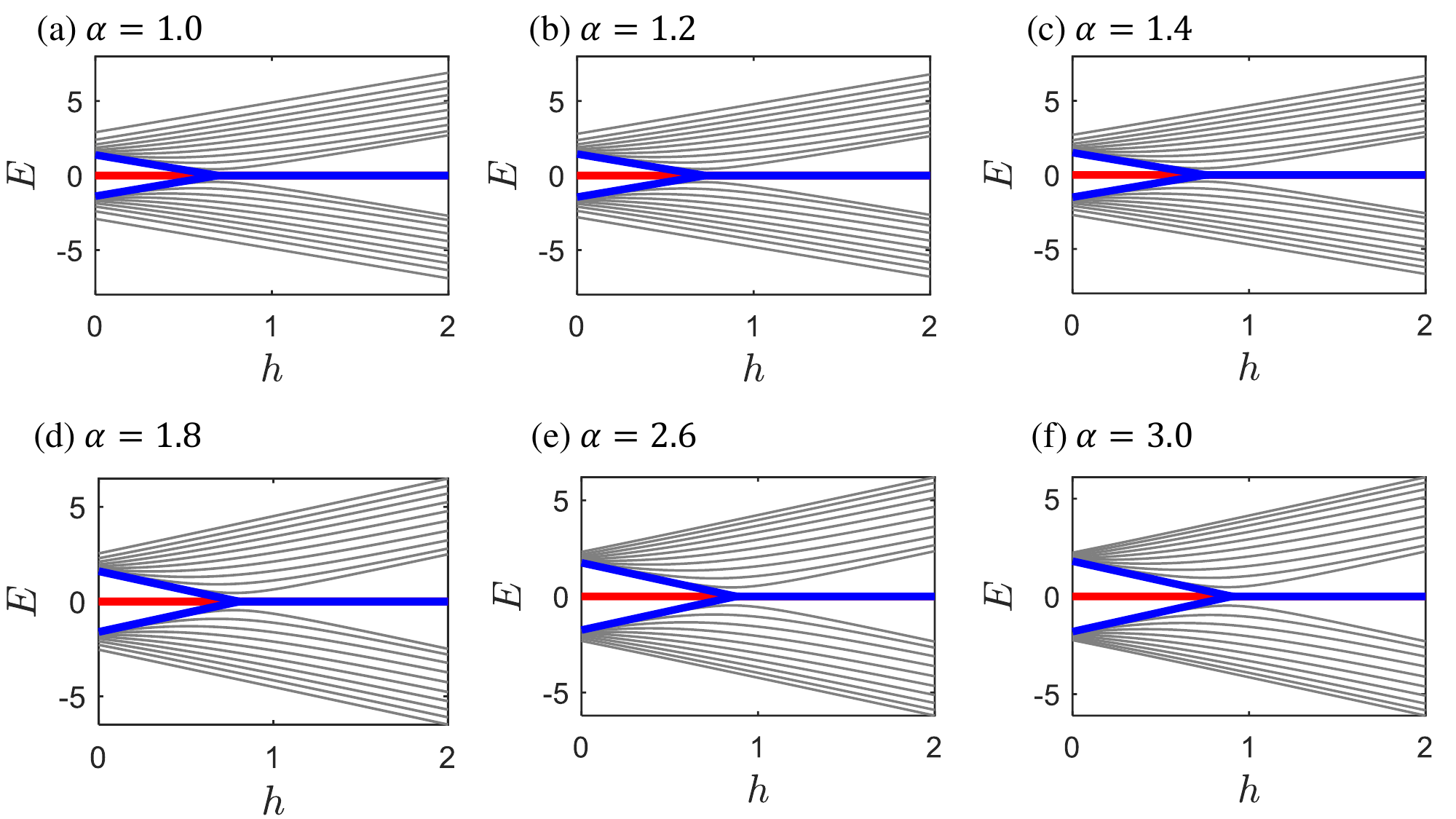}
\caption{(Color online) Energy spectrum for (a) $\alpha=1.0$ (b) $\alpha=1.2$ (c) $\alpha=1.4$ (d) $\alpha=1.8$ (e) $\alpha=2.6$ (f)$\alpha=3.0$ ,and fixed site $L=240$ as a function of parameter $h$.}
\label{fig6}
\end{figure}

Similar to the main text, the energy spectrum as a function of the parameter $h$ for different values of $\alpha$ (a) $\alpha=1.0$, (b) $\alpha=1.2$, (c) $\alpha=1.4$, (d) $\alpha=1.8$, (e) $\alpha=2.6$, and (f) $\alpha=3.0$, with a fixed system size of $L=240$, is illustrated in Fig. 6. It is evident that there is only a pair of zero-energy edge modes in the energy spectrum when $h<h_{c}$, and two pairs of zero-energy edge modes emerge when $h>h_{c}$ (see Fig.~\ref{fig6}).

\begin{figure}
\includegraphics[width=0.8\textwidth]{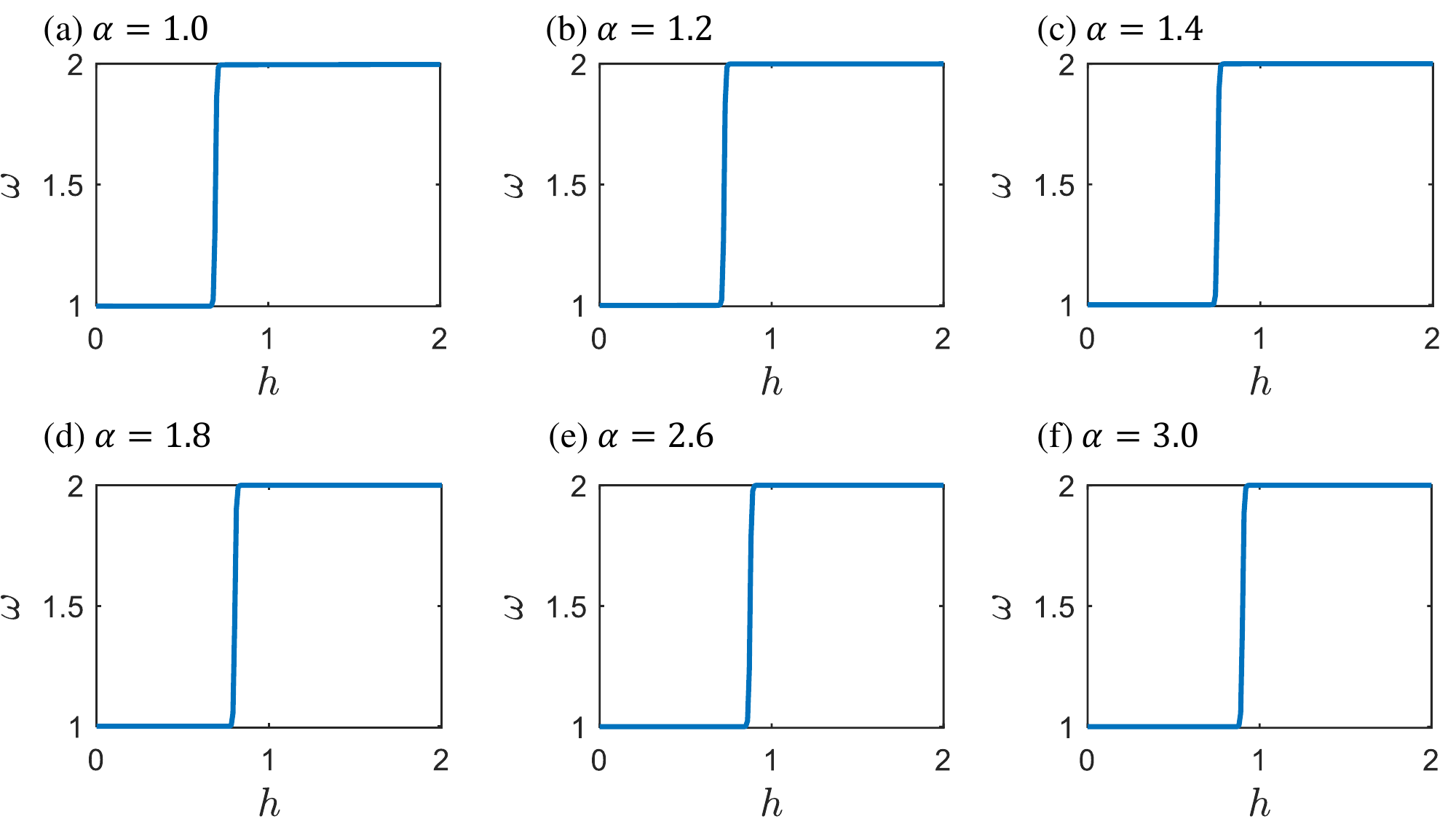}
\caption{(Color online) Winding number  for (a) $\alpha=1.0$ (b) $\alpha=1.2$ (c) $\alpha=1.4$ (d) $\alpha=1.8$ (e) $\alpha=2.6$ (f) $\alpha=3.0$ , and fixed site $L=240$ as a function of parameter $h$.}
\label{fig7}
\end{figure}

On the other hand, the winding number for various values of $\alpha$: (a) $\alpha=1.0$, (b) $\alpha=1.2$, (c) $\alpha=1.4$, (d) $\alpha=1.8$, (e) $\alpha=2.6$, and (f) $\alpha=3.0$, is plotted in Fig.~\ref{fig7} as a function of the parameter $h$ for the system size $L=240$. It is evident that regardless of $\alpha$, the TSC phases of different winding numbers remain stable. Moreover, before and after the phase transition point $h_{c}$, the winding number changes from $1$ to $2$, indicating that the number of edge modes changes from one to two pairs, which is consistent with the observations of the energy spectrum.

\section{FIDELITY SUSCEPTIBILITY AND QUANTUM CRITICAL POINT FOR OTHER INTERACTION POWERS}
\label{sec:appC}
\begin{figure}
\includegraphics[width=0.8\textwidth]{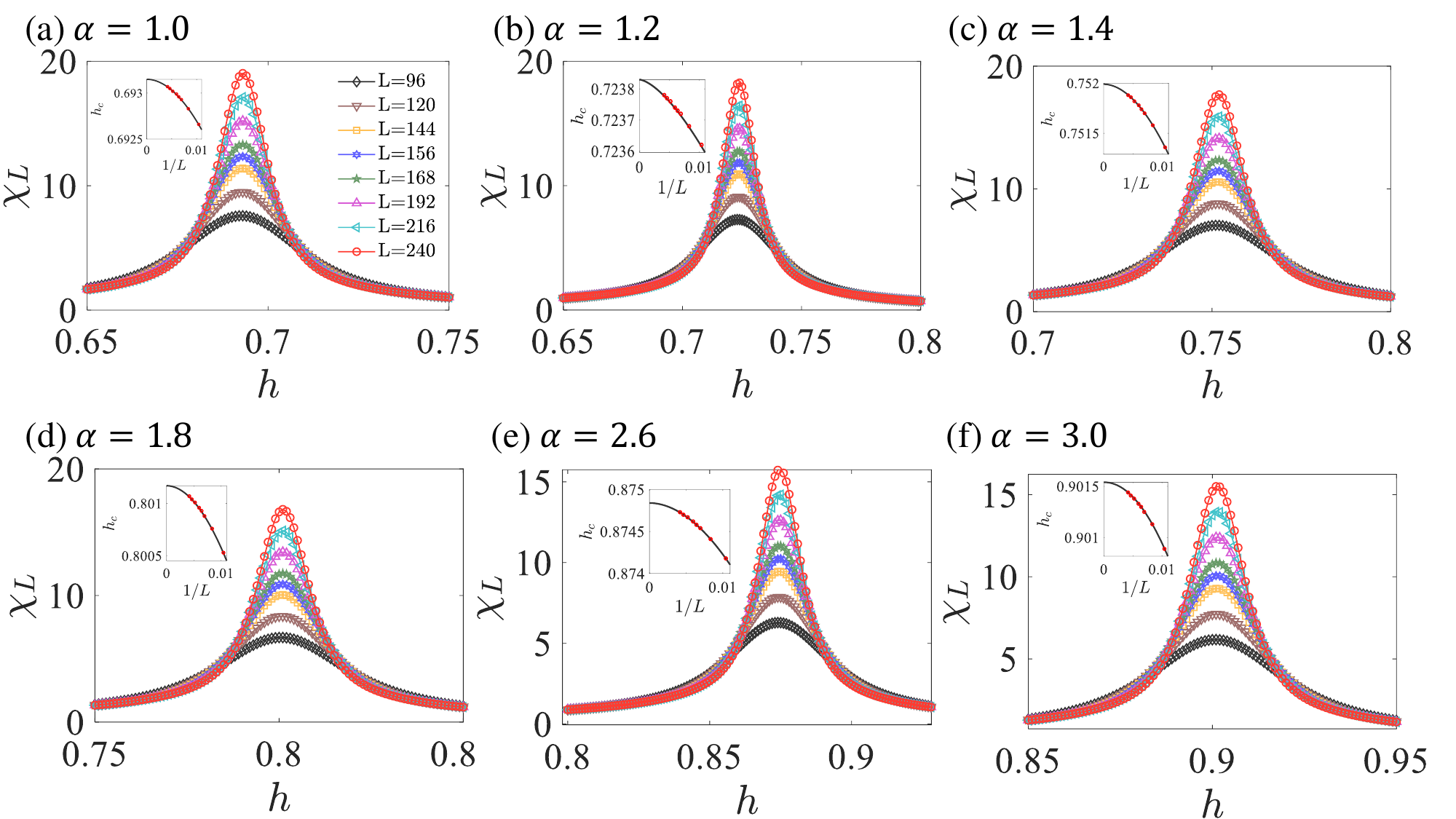}
\caption{(Color online) The fidelity susceptibility per site $\chi_{L}$ is plotted as a function of the parameter $h$ for various values for  (a)$\alpha=1.0$ (b)$\alpha=1.2$ (c)$\alpha=1.4$ (d)$\alpha=1.8$ (e)$\alpha=2.6$ (f)$\alpha=3.0$. The data includes various lattice sizes   $L=96,120,144,156,168,192,216,240$. Additionally, the insert presents a finite-size scaling analysis of the critical point for each $\alpha$ value, as a function of the lattice size $L$. To determine the critical point in the thermodynamic limit, we utilize the fitting formula $h_{c} (L) = h_{c}^{*} + aL^{-1/\nu}$, where $h_{c}^{*}$ represents the critical point in thermodynamic limit. }
\label{fig8}
\end{figure}
In this section, we provide additional data to illustrate the fidelity susceptibility $\chi_L$ and the fitting of the quantum critical point for other interaction powers $\alpha$.

Similar to the main text, on one hand, the fidelity susceptibility per site is presented in Fig.~\ref{fig8} for various values of $\alpha$: (a) $\alpha=1.0$, (b) $\alpha=1.2$, (c) $\alpha=1.4$, (d) $\alpha=1.8$, (e) $\alpha=2.6$, and (f) $\alpha=3.0$. The data is plotted as a function of the parameter $h$ for lattice sizes of $L=96,120,144,156,168,192,216,240$ sites.

On the other hand, the insert in Fig.~\ref{fig8} illustrates the finite-size scaling analysis for the critical point, considering the same range of $\alpha$ values and as a function of the lattice size $L$. Utilizing the fitting formula $h_{c}(L) = h_{c}^{*} + aL^{-1/\nu}$, we determine the critical point in the thermodynamic limit, denoted as $h_{c}^{*}$, and find that the quantum critical point shifts to higher $h_{c}^{*}$ values as $\alpha$ increases.

\section{DATA COLLAPSES AND CRITICAL EXPONENTS FOR OTHER INTERACTION POWERS}
\label{sec:appD}
\begin{figure}
\includegraphics[width=0.8\textwidth]{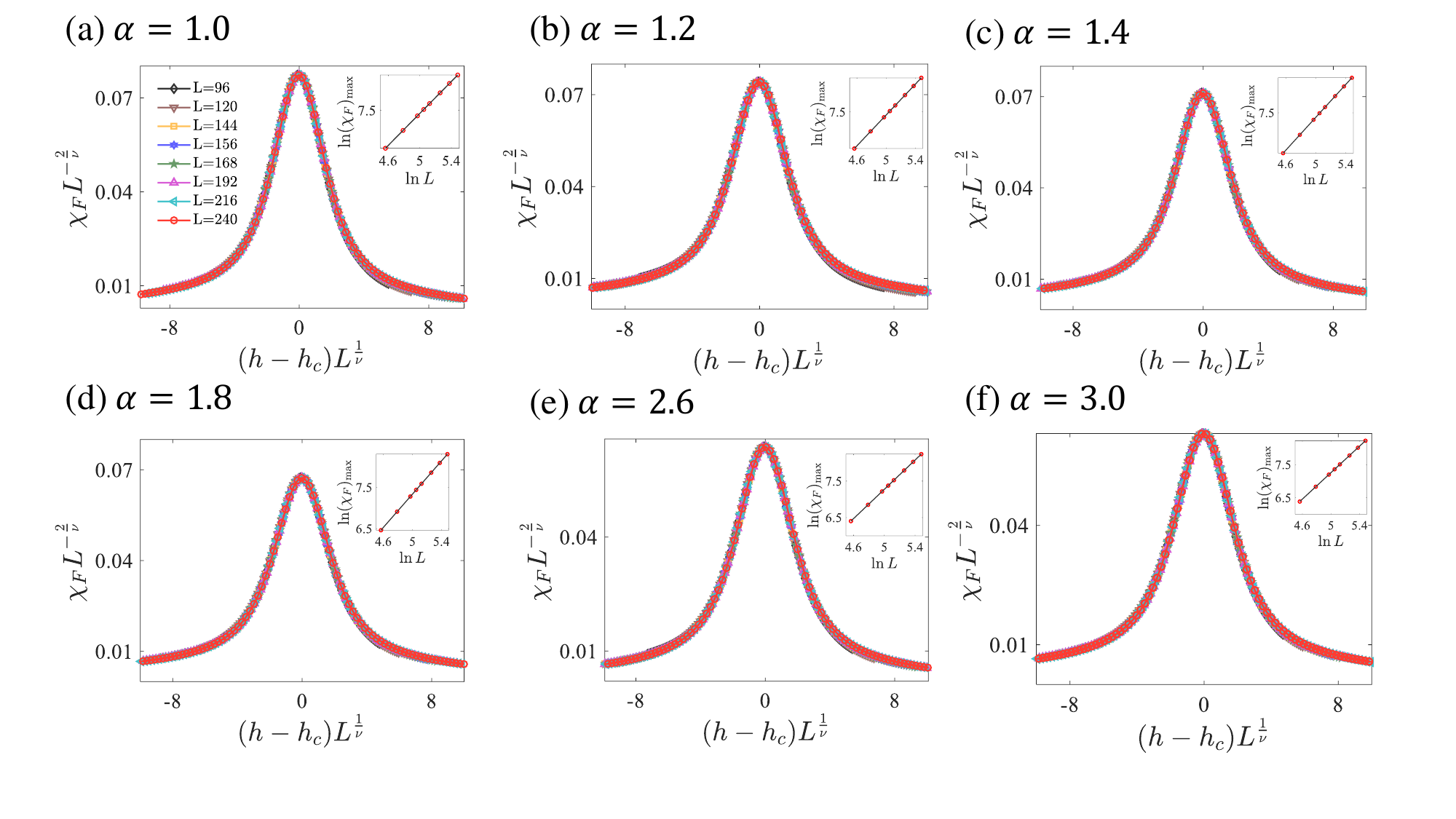}
\caption{(Color online) Data collapse analysis of fidelity susceptibility  $\chi_F$ involved investigating various scenarios under different LR power exponents $\alpha$, including (a) $\alpha=1.0$, (b) $\alpha=1.2$, (c) $\alpha=1.4$, (d) $\alpha=1.8$, (e) $\alpha=2.6$, and (f) $\alpha=3.0$. Additionally, we performed finite-size scaling for the maximum fidelity susceptibility, as illustrated in the insert. To obtain the critical exponent $\mu$, we employed the fitting formula $\chi_{F}(h_{c}(L))=L^{\mu}(c+dL^{-1})$ for our analysis.} 
\label{fig10}
\end{figure}

In this section, we provide additional data to extrapolate critical exponents and demonstrate the accuracy of the estimated critical point and critical exponent for other interaction powers $\alpha$.

Similar to the main text, on one hand, data collapse of the fidelity susceptibility $\chi_F$ for (a) $\alpha=1.0$, (b) $\alpha=1.2$, (c) $\alpha=1.4$, (d) $\alpha=1.8$, (e) $\alpha=2.6$, and (f) $\alpha=3.0$, with different lattice sizes $L$, is shown in \Fig{fig10}. All fidelity susceptibilities for different $L$ collapse into a single curve, indicating the accuracy of the estimated critical point and critical exponent. On the other hand, to determine the critical adiabatic dimension $\mu$, we also performed finite-size scaling for the maximum of the fidelity susceptibility for (a) $\alpha=1.0$, (b) $\alpha=1.2$, (c) $\alpha=1.4$, (d) $\alpha=1.8$, (e) $\alpha=2.6$, and (f) $\alpha=3.0$ as a function of lattice size $L$, as shown in the insert in Fig.~\ref{fig10}. We used the fitting formula $\chi_{F}(h_{c}(L))=L^{\mu}(c+dL^{-1})$ to obtain the critical adiabatic dimension $\mu$. The results exhibit that the critical adiabatic dimension $\mu$ as a function of $1/\alpha$ remains relatively constant and approaches the critical exponent of the Kitaev chain in the SR limit.

\section{WAVE FUNCTION DISTRIBUTION AND EDGE MODE MASS AT THE CRITICAL POINT FOR OTHER
INTERACTION POWERS}
\label{sec:AppE}
The Hamiltonian \eqref{E1} under OBCs can be diagonalized as $H=\sum_{n=1}^{L} \Lambda_{n}\eta^{\dagger}_{n}\eta_{n}$ through a canonical Bogoliubov transformation by introducing the fermionic operators $\eta_{n}$ and $\eta^{\dagger}_{n}$,

\begin{equation}
\eta_{n}=\sum^{N}_{j}(u^{*}_{n,j}c_{j}+v_{n,j}c^{\dagger}_{j}),~\eta^{\dagger}_{n}=\sum^{N}_{j}(u_{n,j}c^{\dagger}_{j}+v^{*}_{n,j}c_{j}),
\end{equation}
where $u_{n,j}$ and $v_{n,j}$ denote the two components of the wave function at site $j$, $n$ is the energy band index, and $\Lambda_{n}$ represents the eigenstate energy. The Schr\"{o}dinger equation $H\ket{\Psi_{n}}=E_{n}\ket{\Psi_{n}}$ can be written as

\begin{equation}
\begin{pmatrix}
A & B \\
-B^{*} & -A^{T}
\end{pmatrix}
\begin{pmatrix}
u_{n,i} \\
v_{n,i}^{*}
\end{pmatrix}=
E_{n}
\begin{pmatrix}
u_{n,i} \\
v_{n,i}^{*}
\end{pmatrix},
\end{equation}
where $A(B)$ is a $N\times N$ symmetric (antisymmetric) matrix. The wave function can be computed as $\lvert \Psi_{n,j}\rvert^{2}=\lvert u_{n,j}\rvert^{2}+\lvert v_{n,j}\rvert^{2}$.

\begin{figure}
\includegraphics[width=0.8\textwidth]{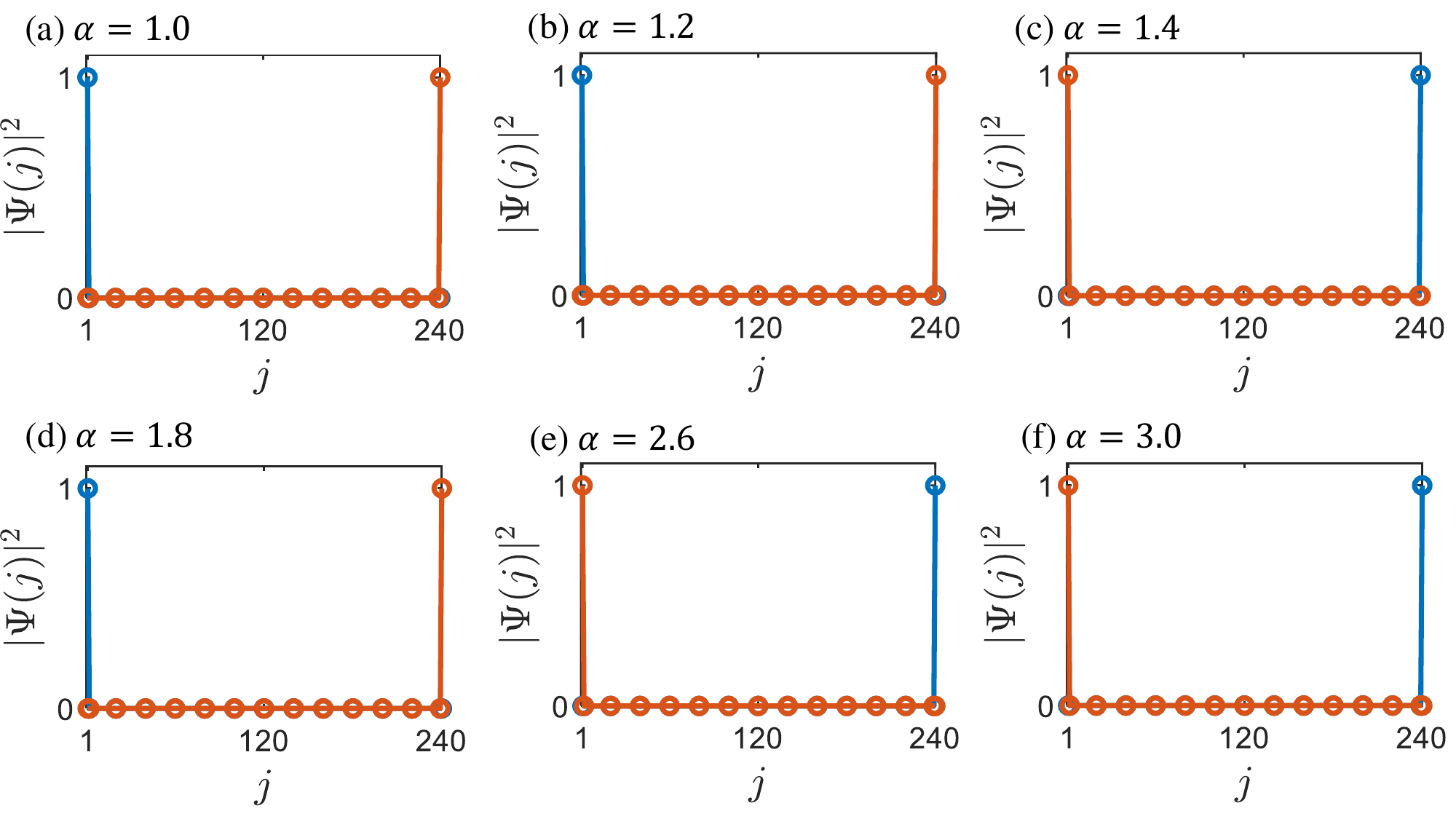}
\caption{(Color online) Wave function distribution at the critical point for (a) $\alpha=1.0$, (b) $\alpha=1.2$, (c) $\alpha=1.4$, (d) $\alpha=1.8$, (e) $\alpha=2.6$, (f)$\alpha=3.0$ ,and fixed system sizes $L=240$.} 
\label{fig11}
\end{figure}

As illustrated in the main text, the wave function distributions at the critical points for different values of $\alpha$: (a) $\alpha=1.0$, (b) $\alpha=1.2$, (c) $\alpha=1.4$, (d) $\alpha=1.8$, (e) $\alpha=2.6$, and (f) $\alpha=3.0$, with a fixed lattice size of $L=240$, are depicted in \Fig{fig11}. We observe that regardless of the strength of the LR interaction, the wave function distribution at the critical point consistently exhibits prominent peaks at the boundary. This observation underscores the robustness of the critical edge mode even under substantially strong LR interactions.

On the other hand, similar to the long-range gapped topological phase, to explore whether long-range interaction may turn the massless edge mode into a massive one, we define the mass gap at finite size $L$ as
\begin{equation}\label{lambda0}
\Lambda_{0}=\min_{n} \Lambda_{n}.
\end{equation}

\begin{figure}
\includegraphics[width=0.8\textwidth]{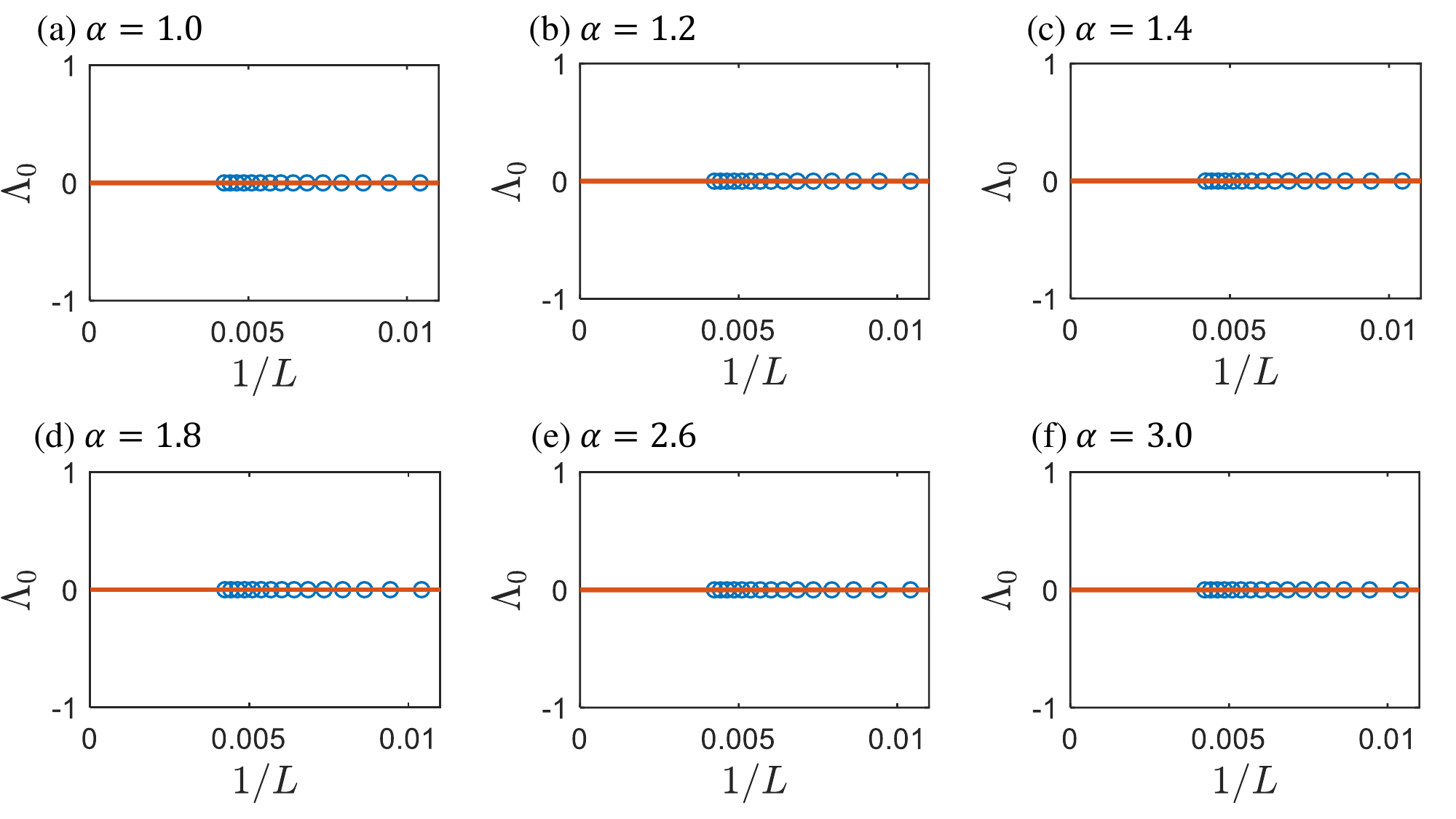}
\caption{ (Color online) The edge mode mass as a function of inverse system sizes $1/L$ for (a) $\alpha=1.0$, (b) $\alpha=1.2$, (c) $\alpha=1.4$, (d) $\alpha=1.8$, (e) $\alpha=2.6$, (f) $\alpha=3.0$.} 
\label{fig9}
\end{figure}

As in the main text, the edge mode mass at the critical point as a function of the inverse system sizes $1/L$ for (a) $\alpha=1.0$, (b) $\alpha=1.2$, (c) $\alpha=1.4$, (d) $\alpha=1.8$, (e) $\alpha=2.6$, and (f) $\alpha=3.0$, are shown in \Fig{fig9}. We observe that the edge mode mass remains zero at the thermodynamic limit even with substantial small $\alpha$, which is entirely different from the emergence of massive edge modes in gapped topological phases.

\end{document}